\documentclass[journal]{IEEEtran}

\ifCLASSINFOpdf
\else
   \usepackage[dvips]{graphicx}
\fi
\usepackage{url}

\usepackage{graphicx}
\usepackage[T1]{fontenc} % add special characters (e.g., umlaute)
\usepackage[utf8]{inputenc} % set utf-8 as default input encoding
\usepackage{amsmath,cite,url}
\usepackage{graphicx}
\usepackage{color}
\usepackage{tikz}
\usepackage{subcaption}
\usepackage{graphicx}
\usetikzlibrary{positioning, shapes.geometric, arrows.meta}
\usepackage{multirow,booktabs}
\begin{document}

\title{Multi-label Cross-lingual automatic music genre classification from lyrics with Sentence BERT}

% Note: Please do NOT use \thanks or a \footnote in any of the author markup

% Single address
% To use with only one author or several with the same address
% ---------------
%\oneauthor
% {Names should be omitted for double-blind reviewing}
% {Affiliations should be omitted for double-blind reviewing}

% Two addresses
% --------------
%\twoauthors
%  {First author} {School \\ Department}
%  {Second author} {Company \\ Address}

% Three addresses
% --------------\input{ISMIR2021_paper.tex}

\author{Tiago F. Tavares, Fabio J. Ayres
\thanks{Submitted for review on January 6th, 2025.}
\thanks{T. Tavares and F. Ayres are with INSPER Institute of Teaching and Research. R. Quatá, 300, São Paulo-SP, Brazil, 04546-042. \{tiagoft1,fjayres\}@insper.edu.br.}}

\markboth{Journal of \LaTeX\ Class Files, Vol. 14, No. 8, August 2015}
{Shell \MakeLowercase{\textit{et al.}}: Bare Demo of IEEEtran.cls for IEEE Journals}
\maketitle

%

%

% - Classificador one-vs-all treina lingua 1 testa lingua 2 mede acuracia e compara com silhouette lingua 1 genero atual vs lingua 2 genero atual

% - UMAP/PCA de todos os embeddings com coloração lingua/genero, para verificar sobreposição

% - Questionamentos futuros: seria a diferença devida ao modelo multilingual, ou a diferenças culturais reais entre linguas.

\begin{abstract}
Music genres are shaped by both the stylistic features of songs and the cultural preferences of artists' audiences. Automatic classification of music genres using lyrics can be useful in several applications such as recommendation systems, playlist creation, and library organization. We present a multi-label, cross-lingual genre classification system based on multilingual sentence embeddings generated by sBERT. Using a bilingual Portuguese-English dataset with eight overlapping genres, we demonstrate the system’s ability to train on lyrics in one language and predict genres in another. Our approach outperforms the baseline approach of translating lyrics and using a bag-of-words representation, improving the genrewise average F1-Score from $0.35$ to $0.69$. The classifier uses a one-vs-all architecture, enabling it to assign multiple genre labels to a single lyric. Experimental results reveal that dataset centralization notably improves cross-lingual performance. This approach offers a scalable solution for genre classification across underrepresented languages and cultural domains, advancing the capabilities of music information retrieval systems.
\end{abstract}
\section{Introduction}\label{sec:introduction}

Music genres are labels that reflect both the style and the audience of songs and artists~\cite{ke_nie_2022_7342846}. As such, they can be used in music recommendation, curation, playlist generation, and listening behavior analysis~\cite{Minz2021}. Most automatic music genre classification methods rely on analysis of audio signals~\cite{Shajin2022}, but there is significant interest in using music lyrics as inputs for this task~\cite{mayerl2022} as they can be important sources of information related to compositional and stylistic choices for a song~\cite{Ogihara2018,Fang2018}.

Automatic music genre identifiers using music lyrics have explored a myriad of different models and feature-engineering strategies~\cite{Mayer2011,Cyril2008,fell2014-lyrics,Mayer2008RhymeAS,mayer2010,Neumayer2007,Ying2012,Tsaptsinos2017MusicGC,mayerl2022}. However, datasets have been linked to specific languages -- mostly English, but also with initiatives for Bangla~\cite{Hamza2023}, Portuguese and Spanish~\cite{Howard2011}, and Nordic songs~\cite{Lima2014}. This is due to the language-specificity of most features used for genre classification (either  extracted by \emph{deep learning} techniques~\cite{Tsaptsinos2017MusicGC} or handcrafted~\cite{mayerl2022}), which makes it necessary to label a whole dataset to train and evaluate models in each language.

A significant part of recent work in Natural Language Processing has been devoted to solve the problem of translation by finding a middlepoint embedding that encodes the meaning of sentences in a given language so that they can be decoded to any other language in a further step~\cite{NIPS2014_a14ac55a,bahdanau2016neuralmachinetranslationjointly}. More recently, there has been work in finding multilingual sentence embeddings, namely sentence-BERT, or sBERT, that were specifically trained to map sentence translations in multiple languages to similar vectors without being linked to a translation downstream task~\cite{reimers-2019-sentence-bert}. These multiligual representations can foster cross-lingual lyric-based music genre identifiers, that is, machines trained to identify genres in one language and can identify similar genres in another language.

In this work, we exploit multilingual embeddings to build a lyrics-based genre classification system that operates across languages, that is, that can be trained in lyrics from one language and then applied in lyrics from another language. The proposed approach outperforms the baseline technique of translating~\cite{klein2018opennmt} the lyrics in the test set and applying a traditional bag-of-words representation. Also, we verify that centralizing the training and test sets, which mitigates the domain shift in sBERT embeddings caused by changing languages, can increase the system performance while still not requiring labeling the test set.

Our proposal consists of a multi-label classifier, that is, a system that can attribute zero or more labels to each lyric. This is an important attribute because our data contains lyrics that can be labeled as more than one genre (e.g., ``Pop'' and ``Rock''). This was attained by using a different one-vs-all classifier for each genre in the training set.

%We use the available data to identify eight music genres that are highly present in both languages, and use them for cross-lingual (that is, train in one language and test in another) experiments.

%Our results indicate that using a cross-lingual training with multilingual embeddings~\cite{reimers-2019-sentence-bert} is an effective technique, but is harmed by cultural differences within each genre.

% In the experimental setup, we extract lyrics embeddings used a pre-trained sBERT~\cite{reimers-2019-sentence-bert}, and proceed to a cross-lingual train-test procedure (that is, train in lyrics from one language and test in lyrics from another language) using a Support Vector Machine classifier over these eight genres.

The proposed method is further discussed next.

\section{Proposed method}
\label{sec:proposed}
The proposed method, as depicted in Figure~\ref{fig:proposed}, uses a pre-trained sBERT to generate embeddings for each lyric in the dataset. The pre-trained sBERT model used was the \textsc{paraphrase-multilingual-mpnet-base-v2}, as available in the \textsc{Sentence Transformers} library, chosen because it has the greatest performance among the available multilingual models. This model has a contetx window of 128 tokens, thus songs were broken into sentences (using cues such as punctuations) and the average of all sentence embeddings were used as the embedding for that particular song. 

\begin{figure}[h!]
\centering
\begin{tikzpicture}[
    node distance=0.2cm and 0.5cm, % distance between nodes
    every node/.style={font=\sffamily}, % general font for all nodes
    process/.style={rectangle, draw, rounded corners, align=center}, % process style
    data/.style={align=center}, % data style
    output/.style={align=center}, % output style
    baselineprocess/.style={rectangle, draw, rounded corners, dashed, align=center},
    arrow/.style={-Latex, thick}, % arrow style
    baselinearrow/.style={-Latex, thick, dashed}
]

% Text input
\node[data] (input) {Lyrics};

% sBERT process
\node[process, right=of input] (sbert) {sBERT};

\node[baselineprocess, below=of sbert] (bow) {BoW};

% SVM process
\node[process, right=of sbert] (clf) {SVM};
\node[baselineprocess, below=of clf] (baseline-clf) {SVM};

% Prediction output
\node[output, right=of clf] (output) {Predictions};

\node[output, yshift=0.2cm, below=of output] (output2) {Baseline\\Predictions};

% Arrows
\draw[arrow] (input) -- (sbert);
\draw[arrow] (sbert) -- (clf);
\draw[arrow] (clf) -- (output);
\draw[baselinearrow] (input) |- (bow);
\draw[baselinearrow] (bow) -- (baseline-clf);
\draw[baselinearrow] (baseline-clf) -- (output2);
\end{tikzpicture}

\caption{Proposed method. Lyrics are first processed by a pre-trained sBERT. This yields embeddings which are further classified by a Support Vector Machine, whose output is a genre prediction. A Bag-of-Words representation based on a TF-IDF vectorizer is used as a baseline for performance comparison.}
\label{fig:proposed}
\end{figure}

The embeddings provided by sBERT are yielded to a Support Vector Machine (SVM) classifier. They were not further scaled because sBERT yields values contained within a $[-1,+1]$ range. The classifier was trained to identify if a lyric is associated with a particular genre, considering the dataset groundtruth.

The embeddings provided by sBERT are multilingual, which means that this system can be trained in genres from one language and then tested in similar genres from another language. However, the observed differences when transiting between languages could be either due to cultural differences between the typical genre themes, or due to a shift in the embeddings domain caused by changing the language. To mitigate that, we centralize the train and test sets by subtracting their average values. Both variants (with and without centralization) were tested.

As a baseline, we use a TF-IDF Bag-of-Words representation. This representation was trained for each case and excluded words tha were too rare ($\text{DF}<0.01$), too common ($\text{DF}>0.3$), were parts of artist names, or words that were musical terms such as ``chorus'', ``intro'', and so on. These exclusions aimed at regularizing the representation, avoiding polarization towards specific words or lyric transcription artifacts.

The experimental setup is further discussed next.

\section{Dataset and experimental setup}
\label{sec:experiments}
We use a dataset scraped from the \url{vagalume.com.br} website~\cite{neisse2021scrappedlyrics} containing lyrics of popular songs in English and in Portuguese, in $79$ different genres\footnote{This dataset was scrapped in 2021 and is available for download on Kaggle \url{http://www.kaggle.com}}. Within this dataset, genres are characteristics of artists, not of songs, that is, a genre is a label identifying each artist's fanbase, thus can be unrelated to historical or musicological perspectives. Also, within this dataset genres are not mutually exclusively, that is, a song can be simultaneously labeled as ``Pop'' and ``Rock''.

We observed that the languages of some lyrics were mislabeled. Because of that, we use the \textsc{langdetect} Python library to identify the language in each lyric. Lyrics whose original label differed from the detected one were discarded.

To identify the suitable musical genres for our experiment, we found the set $G_{\text{pt}}$ of the 20 genres with most songs in Portuguese, and the set $G_{\text{en}}$ containing the 20 genres with the most song in English. Then, we found the set $G = G_{\text{pt}} \cap G_{\text{en}}$. The number of songs in each language for each of the eight genres in $G$ is shown in Table~\ref{tab:dataset_size}.

% \begin{table}[h!]
% \centering
% \begin{tabular}{|c|c|c|} \hline
% Genre & \# PT & \# EN \\ \hline
% Rock & & \\
% Pop & &  \\ \hline
% \end{tabular}
% \label{tab:dataset_size}
% \caption{Item counts for each language and each music genre in the dataset.}
% \end{table}

\begin{table}[h!]
\centering
\begin{tabular}{|l|r|r|r|} \hline
\multirow{2}{*}{Genres} & \multicolumn{3}{|c|}{Counts} \\ \cline{2-4}
& PT & EN & Total \\ \hline
Rock & 11,636 & 63,565 & 75,201 \\
Romantic & 43,712 & 17,908 & 61,620 \\
Pop & 11,843 & 33,464 & 45,307 \\
Gospel & 35,468 & 7,324 & 42,792 \\
Pop/Rock & 11,557 & 26,313 & 37,870 \\
Hip Hop & 4,837 & 20,873 & 25,710 \\
Rap & 6,733 & 17,082 & 23,815 \\
Country & 6,994 & 10,628 & 17,622 \\  \hline
Total & 96,826 & 138,176 & 235,002 \\ \hline
\end{tabular}

\caption{Item counts for each language and each music genre in the dataset. Most genre names were already international, but we freely translated the name ``Romântico'' to ``Romantic'', keeping the English vernacular in all labels.}
\label{tab:dataset_size}
\end{table}

We translated all songs in Portuguese to English and vice-versa using OpenNMT~\cite{klein2018opennmt}. Each translation was considered a different language for experimental purposes, as the translations should carry the cultural aspects of one language, but the language structures of another. Thus, we worked with four different sets: English (EN), Portuguese (PT), English translated from Portuguese (EN $\leftarrow$ PT), and Portuguese translated from English (PT $\leftarrow$ EN).

We trained a one-vs-all classifier for each different genre. For each case, the train and test sets were balanced by downsampling so that the number of ``positive'' items was the same as the number of ``negative'' items. We report the f1-score for each experiment.

We proceeded to a bootstrapping schema to generate a $\pm 2 \sigma$ interval of confidence for each measure. The procedure was conducted as follows:
\begin{enumerate}
\item For each genre and language, lyrics were randomly split into 80\% for training and 20\% for testing.
\item The training and testing sets were reseampled (with replacement, following the bootstrapping experimental setup) for each genre and language by downsampling the most common class, thus ensuring there is the same number of ``positive'' and ``negative'' samples in each set and that each run cointains a slightly different sample following the same distribution.
\item A one-vs-all classifier was trained for each combination of genre, language for training, and language for testing.
\item During training, the hyperparameter $C$ of the SVM was optimized over the values $\{0.01, 0.1, 1, 10\}$ in a 5-fold cross-validation schema aiming to maximize the f1-score in a validation set.
\item The genrewise f1-score was calculated using the best obtained estimator and the test set.
\item The procedure was repeated from step 1 for 10 times, thus generating bootstraped means and standard deviations for the performance metrics.
\end{enumerate}

\section{Results}
\label{sec:results}

We calculate the mean and standard deviation of the f1-score for each triplet of genre, language for training and language for testing. Then, we summarize these results and report the mean and standard deviation f1-score along all genres for each combination of training and testing languages. We first discuss the baseline results, then the results using sBERT.

\subsection{Baseline}
\label{sec:results_baseline}

The results for the Bag-of-Words representation are shown in Table~\ref{tab:results_bow}. Unsurprisingly, results related to training and testing in the same language are far superior to those related to the cross-lingual experiment. This happens because the cross-lingual setting relies solely on cognates and music-related terms, which are insufficient to provide high performance.

\begin{table}[h!]
\scriptsize
\centering
\begin{tabular}{c|cccc}
\toprule
& \multicolumn{4}{c}{Test}  \\
Train & PT & PT $\leftarrow$ EN & EN & EN $\leftarrow$ PT \\
\midrule
PT & 0.76 ± 0.01 & 0.42 ± 0.05 & 0.44 ± 0.09 & 0.31 ± 0.05 \\
PT $\leftarrow$ EN & 0.67 ± 0.01 & 0.42 ± 0.02 & 0.63 ± 0.00 & 0.46 ± 0.01 \\
EN & 0.30 ± 0.06 & 0.47 ± 0.01 & 0.73 ± 0.00 & 0.35 ± 0.04 \\
EN $\leftarrow$ PT & 0.34 ± 0.03 & 0.42 ± 0.02 & 0.46 ± 0.02 & 0.62 ± 0.01 \\
\bottomrule
\end{tabular}
\caption{F1-score ($\mu \pm 2\sigma$) for baseline (BoW) representation. Simple labels ``PT'' and ``EN'' denote experiments where the data was made of lyrics from that language only. Compound labels ``PT~$\leftarrow$~EN'' and ``EN~$\leftarrow$~PT'' denote experiments where the data was constructed by translating the lyrics from one language into another. For instance, ``EN~$\leftarrow$~PT'' marks the use of Portuguese lyrics translated into English before the experiment.}
\label{tab:results_bow}
\end{table}

However, we note that the $\text{PT}\leftarrow\text{EN}$ training lead to an average $0.67$ f1-score for testing in EN. Also, the EN training had a better performance in the $\text{PT}\leftarrow\text{EN}$ testing when compared to $\text{EN}\leftarrow\text{PT}$. These differences indicate that the translation process could be bypassing some words or expressions that are relevant for classification.

\subsection{Results with uncentralized multilingual embeddings}
When usign multilingual embeddings provided by sBERT, the results sensibly change. As shown in Table~\ref{tab:results_sbert}, training in EN and testing in PT lead to an average F1-score increase from $0.30$ to $0.57$, and training in PT with a test in EN had an increase from $0.44$ to $0.68$. However, the results related to testing with translated texts had less pronouced changes, with some cases presenting a performance decrease when compared to the baseline.
\begin{table}[h!]
\scriptsize
\centering
\begin{tabular}{c|cccc}
\toprule
& \multicolumn{4}{c}{Test}  \\
Train & PT & PT $\leftarrow$ EN & EN & EN $\leftarrow$ PT \\
\midrule
PT & 0.77 ± 0.00 & 0.48 ± 0.02 & 0.68 ± 0.02 & 0.32 ± 0.01 \\
PT $\leftarrow$ EN & 0.49 ± 0.01 & 0.63 ± 0.01 & 0.64 ± 0.01 & 0.43 ± 0.01 \\
EN & 0.57 ± 0.03 & 0.47 ± 0.01 & 0.76 ± 0.00 & 0.33 ± 0.01 \\
EN $\leftarrow$ PT & 0.42 ± 0.01 & 0.44 ± 0.01 & 0.34 ± 0.02 & 0.64 ± 0.01 \\
\bottomrule
\end{tabular}
\caption{F1-score ($\mu \pm 2\sigma$) for multilingual embedding (sBERT) representation. Simple labels ``PT'' and ``EN'' denote experiments where the data was made of lyrics from that language only. Compound labels ``PT~$\leftarrow$~EN'' and ``EN~$\leftarrow$~PT'' denote experiments where the data was constructed by translating the lyrics from one language into another. For instance, ``EN~$\leftarrow$~PT'' marks the use of Portuguese lyrics translated into English before the experiment.}
\label{tab:results_sbert}
\end{table}

As discussed in Section~\ref{sec:proposed}, the performance decrease related to the cross-lingual operation can be either due to an inherent change in the meaning of genre labels in different cultural settings, or due to a domain shift in sBERT embeddings caused by changing language. We evaluate this problem by centralizing the train and test sets in each experiment. Results are shown next.

\subsection{Results with centralized multilingual embeddings}
As shown in Table~\ref{tab:results_sbert_norm}, centralizing the train and test sets lead to an increase in performance for the cross-lingual experiments when compared to their uncentralized counterparts. This means that centralizing was effective in mitigating the effects of a domain change. However, there are cultural differences that cannot be overcame by this technique.
\begin{table}[h!]
\scriptsize
\centering
\begin{tabular}{c|cccc}
\toprule
& \multicolumn{4}{c}{Test}  \\
Train & PT & PT $\leftarrow$ EN & EN & EN $\leftarrow$ PT \\
\midrule
PT & 0.78 ± 0.01 & 0.53 ± 0.01 & 0.69 ± 0.01 & 0.43 ± 0.01 \\
PT $\leftarrow$ EN & 0.57 ± 0.01 & 0.64 ± 0.00 & 0.63 ± 0.01 & 0.50 ± 0.01 \\
EN & 0.69 ± 0.01 & 0.55 ± 0.01 & 0.77 ± 0.00 & 0.47 ± 0.01 \\
EN $\leftarrow$ PT & 0.44 ± 0.01 & 0.50 ± 0.01 & 0.47 ± 0.01 & 0.64 ± 0.01 \\
\bottomrule
\end{tabular}
\caption{F1-score ($\mu \pm 2\sigma$) for normalized multilingual embedding (sBERT-norm) representation. Simple labels ``PT'' and ``EN'' denote experiments where the data was made of lyrics from that language only. Compound labels ``PT~$\leftarrow$~EN'' and ``EN~$\leftarrow$~PT'' denote experiments where the data was constructed by translating the lyrics from one language into another. For instance, ``EN~$\leftarrow$~PT'' marks the use of Portuguese lyrics translated into English before the experiment.}
\label{tab:results_sbert_norm}
\end{table}

Finally, we observe that results with untranslated lyrics (either in training or testing) are consistently superior to those related to translated lyrics. Next, we conduct further discussions on these results.

\section{Discussion}
\label{sec:discussions}
% Our work regards cross-lingual music genre classification using lyrics, which is only possible to the extent that music genres use similar themes across different cultures. The results shown here regard differences between lyrics in English and in Portuguese, which means they are tainted by the cultural differences between English-speaking countries (e.g., United States and England) and Portuguese-speaking countries (e.g., Brazil and Portugal). Results in other languages would be binded to the specific cultural differences found between countries speaking these languages.

To some extent, lyrics-based genre classifiers are intrinsically language-specific, as there are genres that are specific to some cultures, such as the Brazilian Forró or the Peruvian Cueca. Moreover, music genres that have spread worldwide, like Rock, Country, and Hip-Hop, typically go through local-specific cultural adaptations and transformations~\cite{ke_nie_2022_7342846}. These transformations reflect language issues, like the use of words, expressions, and rhymes from different languages, and cultural differences, like references to different places, events, and common metaphors.

One important use case for the proposed method is the idea of training a classifier in one language and using it to identify genres in other languages, which can mitigate the effort of obtaining and labeling a large dataset to train machines. Our results indicate that using multilingual embeddings is more effective than translating lyrics in this task. Although this is an effective technique, multilingual embeddings rely on pre-training systems over a large corpora of aligned texts, which could be unavailable for particular languages.

Our results show that the cross-lingual performance can be further increased if the train and test sets are respectively centralized. This requires a corpus of lyrics in the test set's language. However, the corpus can be unlabeled, thus it can be faster to obtain via scrapping.

It is also important to reinforce that the results shown here are bounded to a specific dataset and a general-purpose configuration. Aditional filtering, such as excluding songs composed before a particular year, could insert biases into the system. These biases must be investigated case by case, as the can potentially make the system more adequate to particular applications, such as finding new trends in music lyrics in a cross-lingual setting.

\section{Conclusion}
\label{sec:conclusion}
This work proposes a system for cross-lingual music genre classification using multilingual sentence embeddings provided by Sentence BERT. It presents a scalable solution to a traditionally language-dependent problem. Our approach represents an important step towards overcoming limitations of monolingual genre classifiers, particularly for underrepresented languages. Additionally, the introduction of dataset centralization highlights a practical method to mitigate embedding domain shifts, enhancing cross-lingual performance.

On top of language differences, cultural variations still represent an important challenge for this type of approach. These differences are hard to measure, and results in this field are likely tainted by biases related to the used dataset. The results are also bounded by the discriminative power of the embeddings. Henceforth, another contribution to this problem would be finding embeddings that are more discriminative towards music genres. Both of these aspects represent great challenges that must be tackled to continue the evolution of cross-lingual classification systems.

% Genres in our dataset: 
% {'Gospel/Religioso', 'Country', 'Hip Hop', 'Pop', 'Rap', 'Rock', 'Pop/Rock', 'Romântico'}

% For bibtex users:
\bibliographystyle{ieeetr}
\bibliography{ISMIRtemplate}

% For non bibtex users:
%\begin{thebibliography}{citations}
% \bibitem{Author:17}
% E.~Author and B.~Authour, ``The title of the conference paper,'' in {\em Proc.
% of the Int. Society for Music Information Retrieval Conf.}, (Suzhou, China),
% pp.~111--117, 2017.
%
% \bibitem{Someone:10}
% A.~Someone, B.~Someone, and C.~Someone, ``The title of the journal paper,''
%  {\em Journal of New Music Research}, vol.~A, pp.~111--222, September 2010.
%
% \bibitem{Person:20}
% O.~Person, {\em Title of the Book}.
% \newblock Montr\'{e}al, Canada: McGill-Queen's University Press, 2021.
%
% \bibitem{Person:09}
% F.~Person and S.~Person, ``Title of a chapter this book,'' in {\em A Book
% Containing Delightful Chapters} (A.~G. Editor, ed.), pp.~58--102, Tokyo,
% Japan: The Publisher, 2009.
%
%
%\end{thebibliography}

\end{document}